# RECENT RESULTS FROM THE ALPHA MAGNETIC SPECTROMETER (AMS) ON THE STS-91

Roberto Battiston[1]

[1] *Dipartimento di Fisica dell' Universitá and Sezione INFN, Via Pascoli, 06100 Perugia, Italy*

*AMS Collaboration*

## ABSTRACT

The Alpha Magnetic Spectrometer (AMS) is a state of the art detector for the extraterrestrial study of matter, antimatter and missing matter. During the ten days long STS-91 precursor flight in may 1998 AMS collected nearly 100 millions of Cosmic Rays on Low Earth Orbit, measuring with high accuracy their composition. We present results on the flux of proton, electron, positron and helium particles in the rigidity region 0.1 to 200 $GV$. Analysis of the under cutoff spectra indicates the existence of a new type of belts containing high energy particles and characterized by a dominance of positrons versus electrons.

## INTRODUCTION

The disappearence of the antimatter (Steigmann, 1976, Kolb and Turner, 1983, Peebles, 1993) and the presence at all scales in our universe of a non luminous components of matter (dark matter) (Ellis, 1998, Turner and Wilzek, 1990) are two of the most intriguing misteries in our current understanding of the structure of the universe.

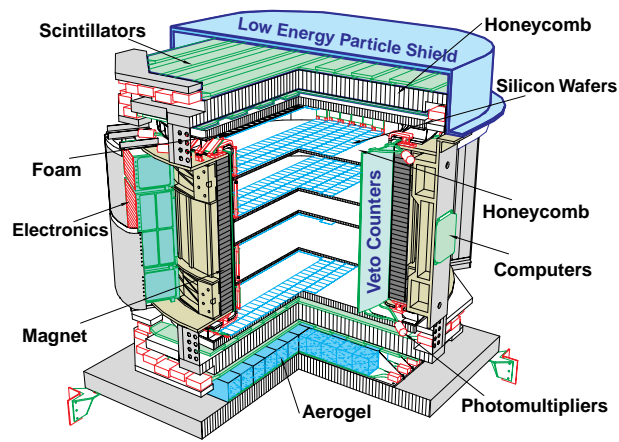

Fig. 1. Configuration of the AMS experiment on the Discovery STS 91 precursor flight, June 2-12, 1998.

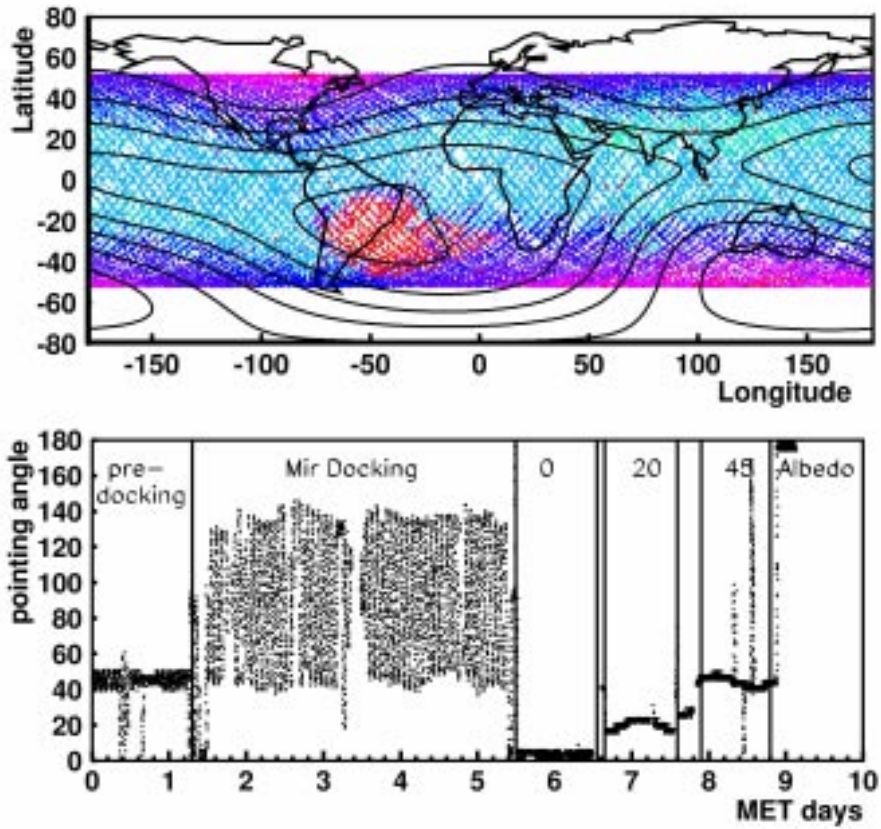

Fig. 2. (a) C.R. rates versus latitude and longitude and (b) shuttle attitudes during the STS91 mission as a function of the Mission Elapsed Time (MET).

To study these problems, a high energy physics experiment, the Alpha Magnetic Spectrometer (AMS) (Ahlen et al., 1994), has been approved and it is scheduled for installation on the International Space Station (ISS)in 2003. Goal of AMS is to perform a three year long measurement, with the highest accuracy, of the composition of Cosmic Rays in the rigidity range 0.1 $GV$ to several $TV$. In preparation for this long duration mission AMS flew a ten days precursor mission on board of the space shuttle Discovery during the STS-91 flight in June 1998. This high statistics measurement of CR in space enabled, for the first time, the systematic study the behaviour of primary CR near Earth in the rigidity interval 0.1 $GV$ to 200 $GV$, at all longitudes and at latitudes up to $\pm51.7^o$. In this paper we present some relevant results obtained by AMS during the precursor mission. We also report the first observation of high energy radiation belts in the near Earth region and on their composition, which shows remarkable differences with previously observed belts of trapped particles around our planet.

## THE AMS EXPERIMENT ON THE STS-91 MISSION

Search of antimatter requires the capability to identify with the highest degree of confidence, the mass of particle traversing the experiment together with the absolute value and the sign of its electric charge.

The AMS configuration flown in 1998 on the Shuttle Discovery (Figure 2) includes a permanent Magnet, Anticounter (ACC) and Time of Flight (ToF) scintillator systems, a large area, high accuracy Silicon

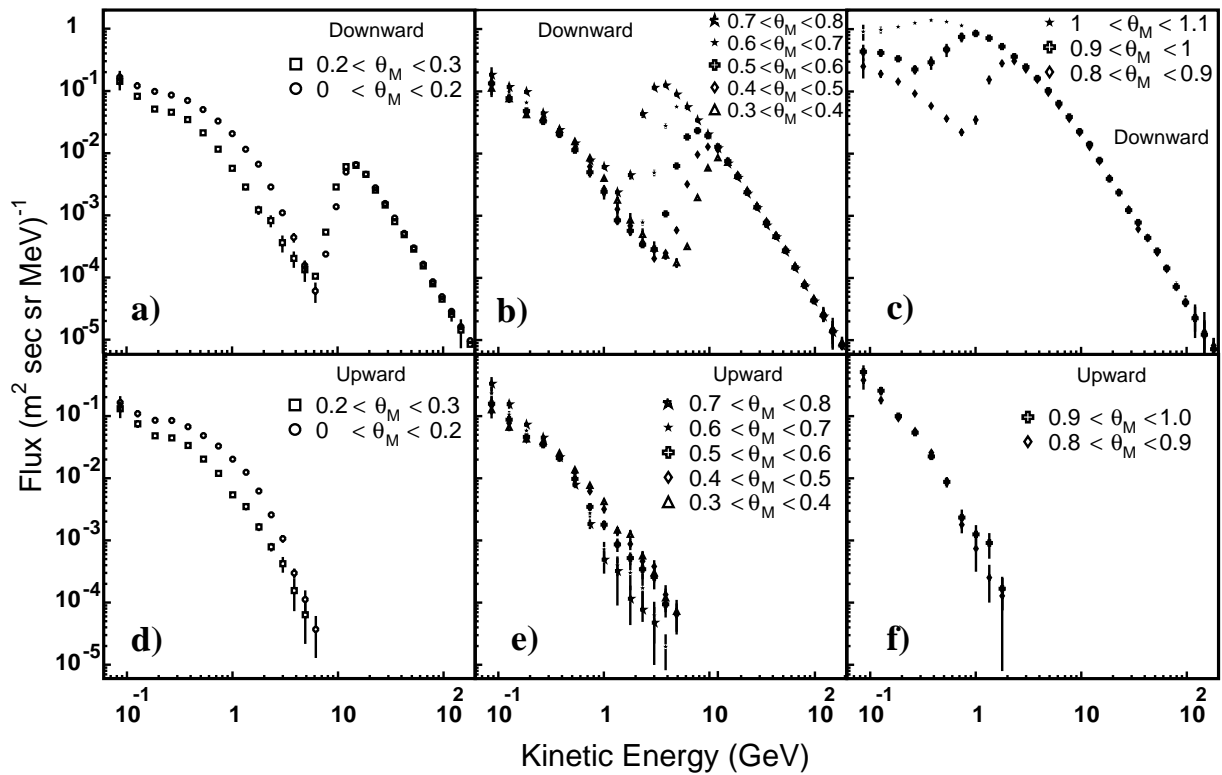

Fig. 3. Proton spectra measured by AMS for different geomagnetic latitude intervals.

Tracker and an Areogel Threshold Cherenkov counter. The magnet is based on recent advancements in permanent magnetic material and technology which makes it possible to use very high grade Nd-Fe-B to build a permanent magnet with $BL^2 = 0.15\ Tm^2$ weighting $\leq 2$ tons. A charged particle traversing the spectrometer triggers the experiment through the ToF system, which measures the particle velocity with a resolution of $\sim 120\ ps$ over a distance of $\sim 1.4\ m$ (Alvisi, 1999).

AMS is the first high energy spectrometer based solely on high precision microstrip silicon detectors. Pattern recognition and tracking is performed using a large area (total surface on ISS $\sim 7\ m^2$, on STS-91 $\sim 2.3\ m^2$ ), multilayer (on ISS 8 layers, on STS-91 6 layers), high accuracy Silicon Tracker (Battiston, 1994, Alcaraz et. al., 2000). On the Space Station mission it will consist of 2300, high purity, double sided silicon wafers, 300 $\mu m$ thick (CSEM, 2000), following a technology developed in Italy by INFN for the Aleph (Batignani et. al., 1989) and L3 (Acciarri et al., 1990) vertex detectors at LEP. The active area of the AMS Silicon Tracker is about an order of magnitude larger than for microstrip silicon detectors currently operating at high energy Colliders.

AMS momentum resolution on the precursor mission was about ($\frac{\Delta p}{p} \sim 7\%$) at 10 $GV$, reaching ($\frac{\Delta p}{p} \sim 100\%$) at about 500 $GV$.

The four ToF scintillators layers and the Silicon Tracker layers measure $\frac{dE}{dx}$, allowing a multiple determination of the absolute value of the particle charge. In addition, two layers of an Aerogel Threshold Counter, provide an additional measurement of the particle velocity.

By combining the various subdetector measurements it is possible to determine the type of particle traversing the magnet and identify interesting particles. The estimated background rejection can reach, for anti-matter searches, one part in 10 billions.

During the period june $2^{nd}$ to june $12^{th}$, 1998, the Shuttle Discovery has performed 154 orbits at an inclination $51.7^o$ and at an altitude varying between 390 to 350 $km$. During this mission AMS collected a total of about 100 Million triggers, at various Shuttle attitudes (Figure 2). In the Figure it can be noticed the period of Shuttle to Mir docking when the Shuttle attitude is rapidly changing with time.

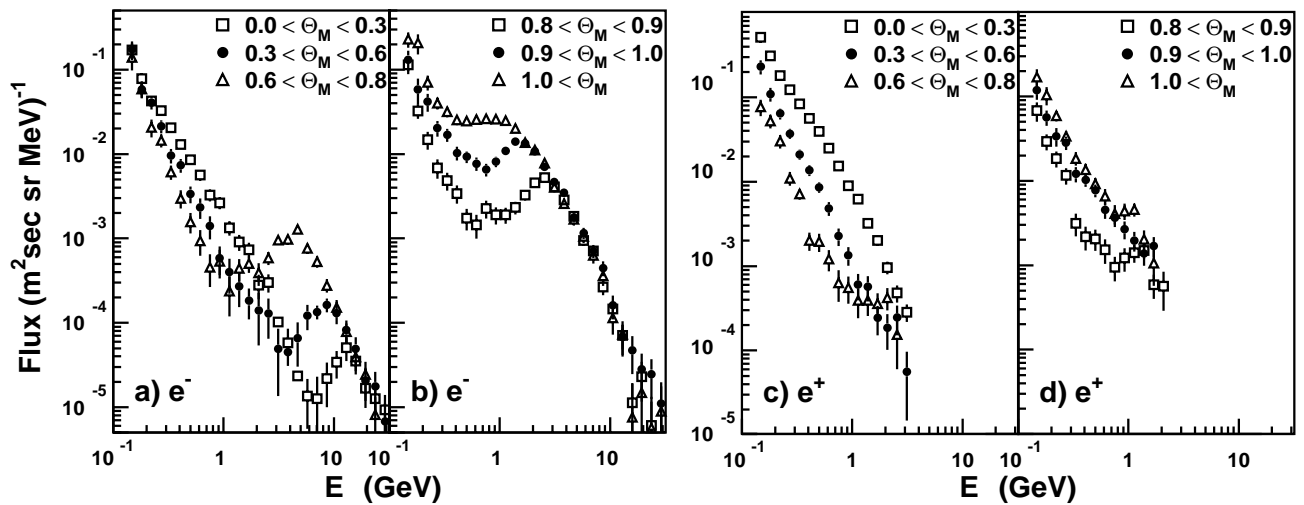

Fig. 4. Electrons and positron spectra measured by AMS during the STS91 flight.

| Proton flux | |
|---|---|
| $\gamma$ | $2.78 \pm 0.009$ (fit) $\pm 0.019$ (syst) |
| $\Phi_o$ | $17.1 \pm 0.15$(fit) $\pm 1.3$(syst) $\pm 1.5(\gamma) GV^{2.78}(m^2 s\ srMeV)^{-1}$ |
| **Helium flux** | |
| $\gamma$ | $2.740 \pm 0.010$(fit)$\pm 0.016$(syst) |
| $\Phi_o$ | $2.52 \pm 0.09$(fit)$\pm 0.13$ (syst) $\pm 0.14(\gamma) GV^{2.78}(m^2 s\ srMeV)^{-1}$ |

Table 1. AMS results on the parametrization of proton and helium primary flux.

Almost all results published so far (AMS Collaboration,2000a,b,c,d) have been obtained with data collected during well defined attitude periods with AMS pointing at $0^o, 20^o$ and $45^o$ with respect to zenith (deep space).

These data are the first high quality CR data collected with a magnetic spectrometer located outside the atmosphere. The measurements cover all geomagnetic longitudes and most latitudes. These data allow a direct and accurate measurement of the CR composition and spectra, as well as a systematic study of the effects of the geomagnetic field.

The measurement of the proton flux as a function of the geomagnetic latitude (Figure 3a-3c), shows that, in addition to the primary CR spectrum there is a substantial second spectrum below the geomagnetic cutoff. This spectrum extends to much lower energy and exhibits some significant latitude dependence close to the equator. These particles cannot come from the deep space since they are on "forbidden" orbits. They are produced in the interaction of the primary CR with the top layers of the atmosphere. A characteristic of the second spectrum is that it is up-down symmetric (Figure 3d-3f).

Second spectra with similar geomagnetic latitude dependence have been detected by AMS in the low energy region also for $e^-$, $e^+$ (AMS Collaboration, 2000b), $D$ (Lamanna, 2000) and, although with lower intensity, $^3He$ (AMS Collaboration, 2000e). Figures 4 and 5 show the results for electrons (Fig. 4a,b), positrons (Fig. 4c,d) and Helium (Fig. 5), respectively. The results on the second spectra are discussed later in this paper.

Adding all data collected above the geomagnetic cutoff it is possible to obtain a precise estimate of the primary CR differential flux. Parametrizing the omnidirectional CR flux as $\Phi(R) = \Phi_o R^\gamma$ ($R$ in $GV$) we obtain the results reported in Table 1.

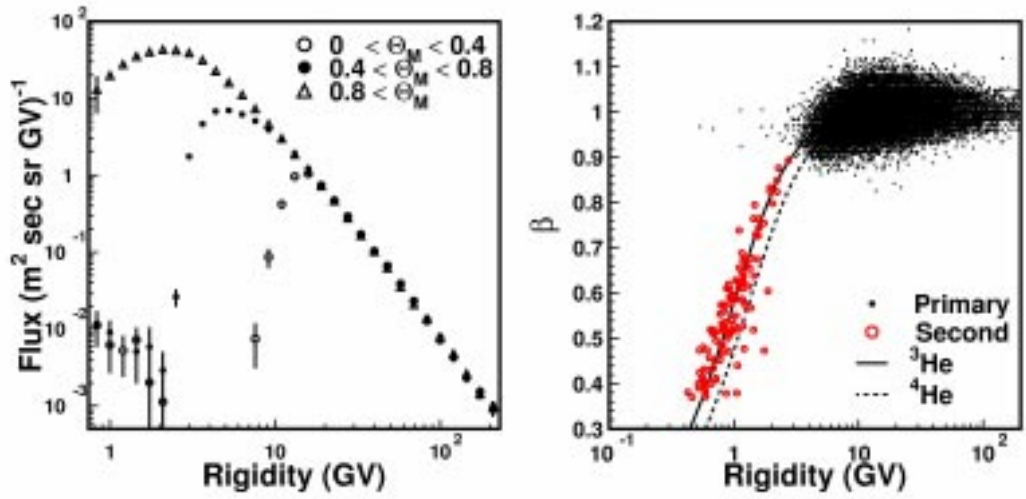

Fig. 5. Helium spectra measured by AMS during the STS91 flight.

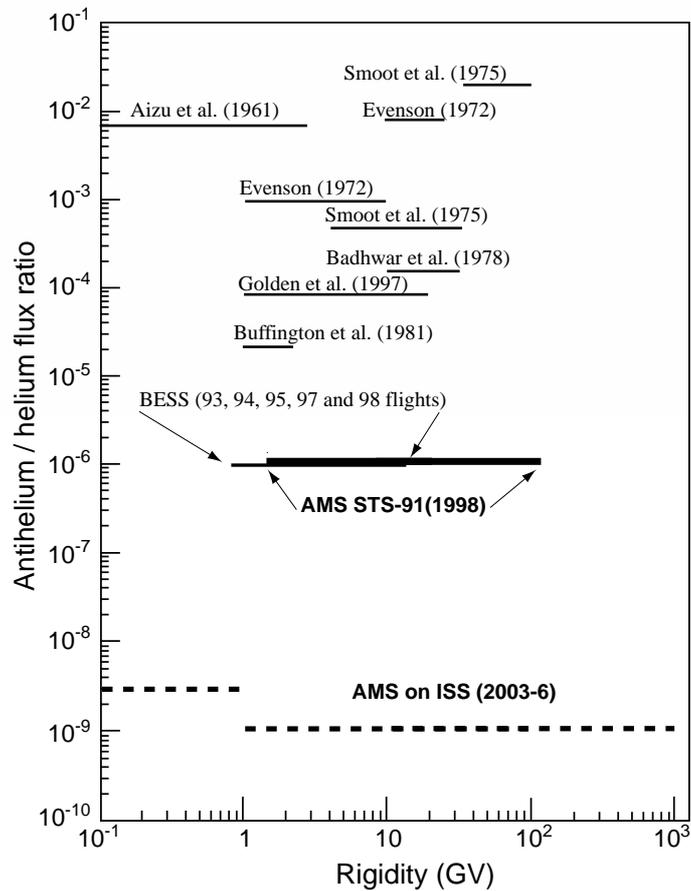

Fig. 6. Compilation of limits on antiHe in C.R..

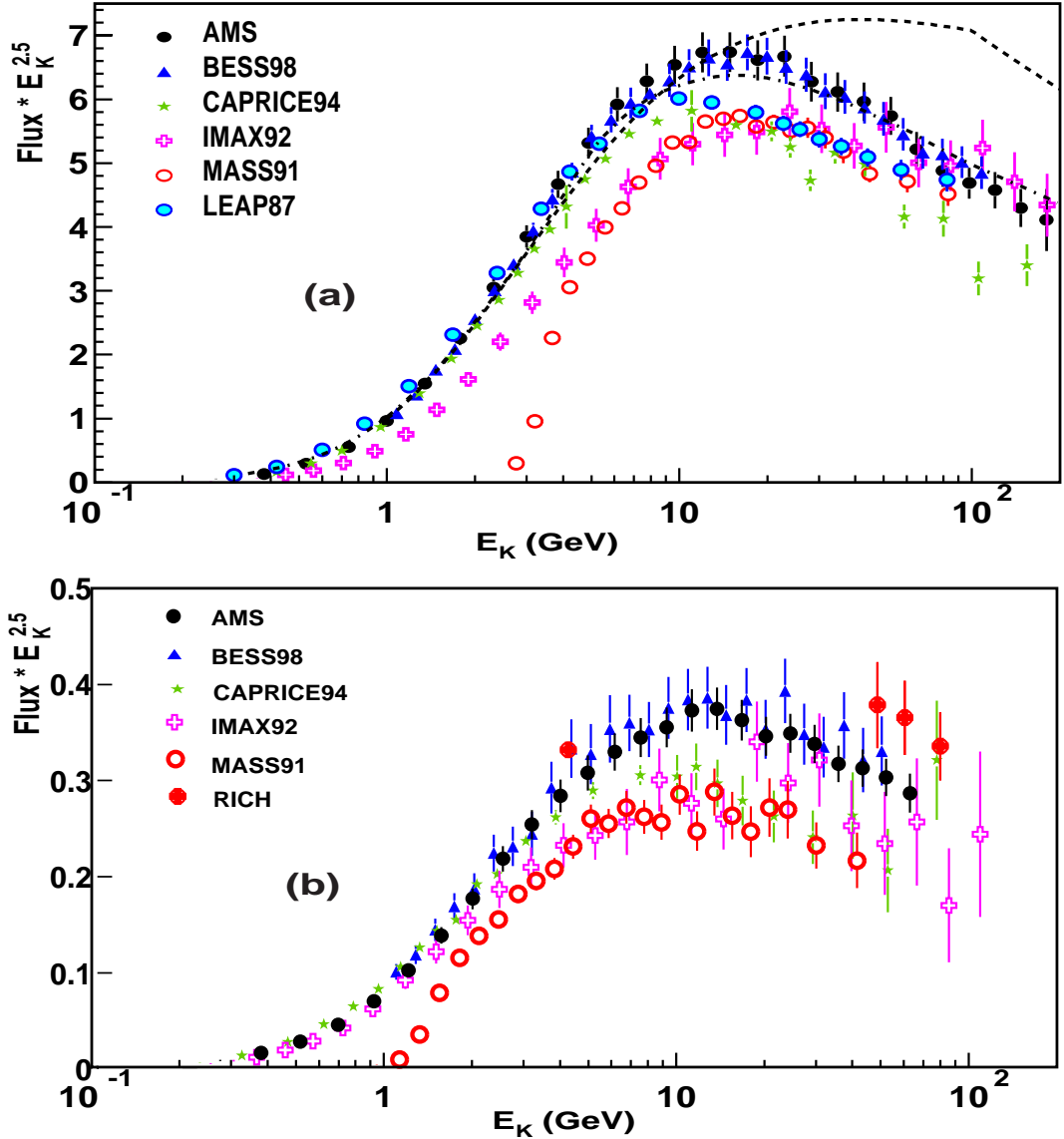

Fig. 7. (a) Primary proton flux measured by AMS and compared with existing balloons measurements. The lines are parametrizations of the primary cosmic rays used in atmospheric $\nu$ flux calculation: dashed line HPPK (Honda, 1995), dot-dashed line Bartol group (Lipari, 2000); (b) primary He flux measured by AMS and compared with existing balloons measurements.

It it interesting to compare AMS measurements of the primary fluxes with previous results obtained with stratospheric balloons (Sanuki, 2000, Boezio, 1999, Menn, 2000, Bellotti, 1999, Seo, 1991, Buckley 1992).

Figure 6 shows the comparison for the proton spectrum, multiplied by $E^{2.5}$. The improved statistical significance and the wider energy interval covered by AMS data is evident: thanks to the improved accuracy obtained with only few days in space, it is possible to clarify the situation resulting from the data published over the last 15 years by the various Collaborations using different implementations of the NASA/NMSU Balloon-Borne Magnet Facility (BBMF) (CAPRICE94, IMAX92, MASS91, LEAP87) and by the Bess Collaboration (BESS98). Similar considerations apply for the comparison of the measurements of Helium primary flux (Figure 7).

Both for proton as well as for Helium, AMS data show a nice agreement with the balloon borne measurement of the Bess Collaboration (BESS98), although our data have smaller statistical errors and extends over a wider energy interval.

Using the large Helium sample, a search for anti-He candidates has also been performed by AMS. Within 2.3 $M$ He events no anti-He candidates have been found, up to a rigidity of 140 $GV$.

Assuming identical He and anti-He spectra we obtain a model dependent upper limit of 1.1 $10^{-6}$ over the rigidity interval 1 to 140 $GV$, which can be compared to previous results (Figure 6).

## OBSERVATION OF HIGH ENERGY PARTICLES BELTS

The trapping of charged particles in the quasi dipolar earth magnetic field is a classical problem, which has been studied in great detail (Walt, 1997) following Van Allen observations (VanAllen, 1958). The basic physical mechanism is well understood. For sufficiently low rigidities, the trapped particles spiralize along orbits defining shells surrounding our planet. These shells are shaped along the magnetic field lines and are roughly symmetric in latitude with respect to the geomagnetic equator (Figure 8).

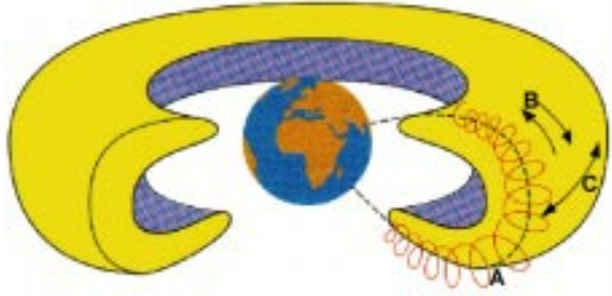

Fig. 8. Motions of a charged particle trapped in the geomagnetic belts. A) gyration B) bouncing C) drift.

The motion of a trapped particle can be separated in three components, the revolution around the guiding center or gyration, the bouncing between mirror points located $\sim$ symmetrically with respect to the geomagnetic equator (magnetic bottle), and a longitudinal drift around the earth. The geometrical locations defined by the orbits of trapped particles are called shells. A shell can be univocally determined by two parameters. For example a pair of variables are L, the distance of the shell at the equator measured in unit of the Earth radius ($R_\oplus$), and $B_{mir}$, the value of the magnetic field at the point where the particles reverse their motion (mirror point)(McIlwain, 1961). Depending on the shell, $B_{mir}$ can be locally very deep in the atmosphere (it can also be below the earth crust). Shells which are characterized by these values of $B_{mir}$ cannot trap the particles, since they are lost within one or few bounces across the magnetic equator. A particle belonging to a shell will remain on the same shell until it is disturbed by (a) interaction with the top layers of the atmosphere or other particles or (b) interaction with electrical or magnetic variable field.

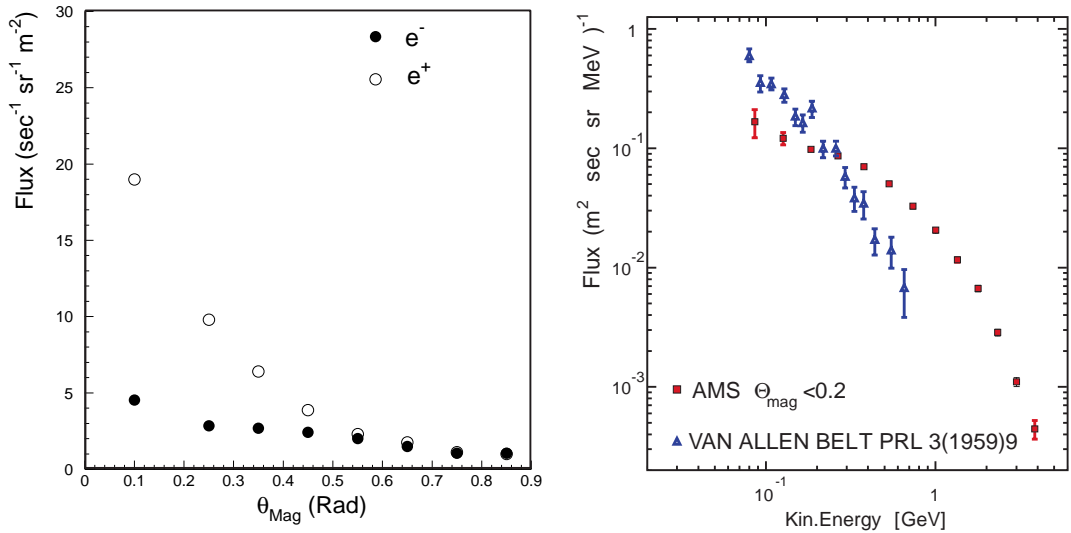

Fig. 9. (a) $\frac{e^+}{e^-}$ ratio inside the belts observed by AMS as a function of the geomagnetic latitude; (b) comparison among a typical Van Allen belt proton spectrum and equatorial AMS belts proton spectrum.

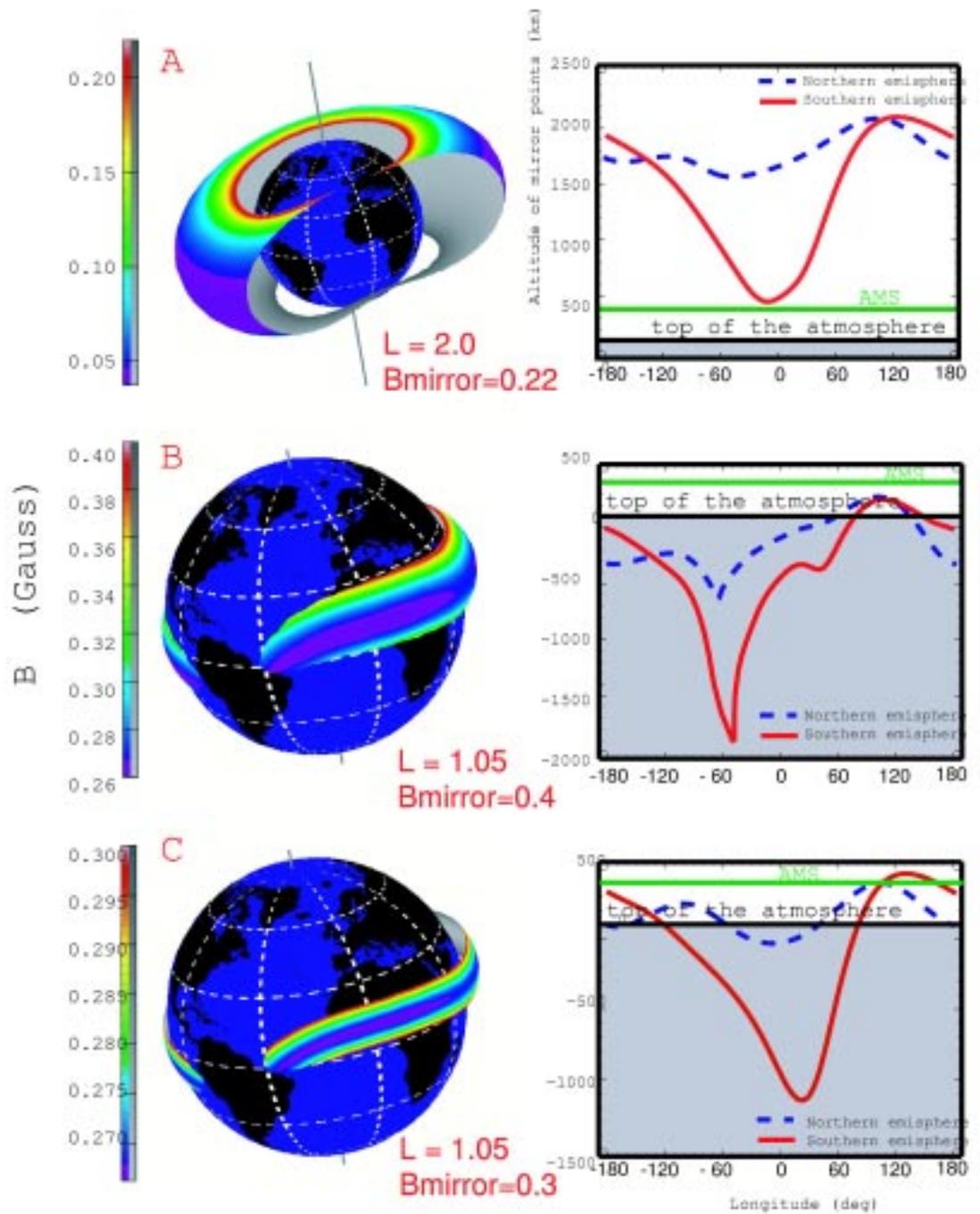

Fig. 10. Van Allen versus AMS belts. (A) Van Allen belts have high L values, $B_{mirr}$ is located mainly above AMS orbits, particles weakly interact with the atmosphere and have lifetimes ranging from days to months. AMS belts have $L < 1.5$, and depending on the value of $B_{mirr}$, their lifetime ranges from fractions of a second (B) to several seconds (C)

| Belt type | Composition | Rigidity [MeV/n] | Filling mechanisms | L | Residence time [d] |
|---|---|---|---|---|---|
| Van Allen (inner) | $p$ | $0.1 - 100$ | $n \rightarrow pe^- \overline{\nu}_e$, | $< 2.5$ | $10 - 1000$ |
| | $e^-$ | $0.01 - 1$ | external belts | | |
| Van Allen (outer) | $e^-$ | $1 - 10$ | solar | $> 2.5$ | $1 - 10$ |
| | $p$ | $0.1 - 1$ | wind | | |
| SAMPEX | $N^{+x}, O^{+x}$, | $10$ | Anomalous | $2$ | $10 - 100$ |
| | $Ne^{+x}$ | $10 - 100$ | CR | | |
| AMS | $p$ | $100 - 1000$ | primary CR | $\leq 1.15$ | $10^{-6} - 10^{-4}$ |
| | $e^-$ | $100 - 1000$ | interacting | | |
| | $e^+$ | $100 - 1000$ | with the | | |
| | $D$ | $100 - 1000$ | atmosphere | | |
| | $^3He$ | $100 - 1000$ | | | |

Table 2. Different types of particle belts around the Earth.

Conversely, primary cosmic rays coming from deep space cannot enter a shell unless their trajectories are disturbed by some interaction with matter or fields. The existence of the shells is the result of the equilibrium between two competing mechanisms: some contributing to fill the shells with new particles and others removing some of the trapped particles.

If the dynamic of the particles trapped in the shell is well understood, the mechanisms contributing to shell stability are much less understood (AGU, 1997). They involve: interaction of high energy CR with the atmosphere creating neutrons which decays in flight, $n \longrightarrow p + e^- + \overline{\nu}_e + 782 KeV$, filling the belts (CRAND mechanism)(Singer, 1958, Kellogg, 1959, Hess, 1959), instabilites due to solar storms, as well as other types of magnetic and electric instabilities. It should be pointed, however, that all mechanisms proposed, are compatible with the observed dominance of protons and electrons in the Van Allen belts.

The shell can be classified by their composition and location. The original Van Allen belts contain only proton and electrons and extend to very large distance from the earth, up to $L \sim 6$. Van Allen belts are divided into inner and outer belts, since there is a dip in the particle flux intensity at about 2.5 $L$. During the last 20 years, there have been reports of the observation of a low flux of trapped ions, mainly $He$ and $O$, with traces of $C$ e $N$, and having energies of a few $MeV/n$ and $L = 3 - 4$. These particles are extracted from the upper layers of the atmosphere during solar storms. More recently, nearly 40 years after the Van Allen discovery, the analysis of SAMPEX data (Cook, 1993) have shown the existence of belts included in the inner Van Allen belts, containing heavier nuclei like $N, O, Ne$ with rigidities of the order of 10 $MeV/n$.

The SAMPEX belts are different from the Van Allen belts mainly because of their composition due to a different filling mechanism, which is likely due to the interaction of the so called Anomalous Cosmic Rays with the Earth atmosphere (Cummings, 1993a,b, Mewaldt, 1997).

The belts observed by AMS are different in composition since they also contain a large fraction of positrons, but also deuterium and $^3He$. These particles have not been observed in the Van Allen or SAMPEX belts. Particularly striking is the abundance of positrons versus electrons (Figure 4), with a ratio exceeding a factor of four in the equatorial region (Figure 9a).

AMS observed shells with $L \leq 1.15$, well below the inner Van Allen belts. In the belts studied by AMS the observed proton spectrum is harder (Figure 9b) than in the case of Van Allen belts. This can be understood since their location is closer to the earth and the particles do experience a stronger trapping field. Another difference with the Van Allen belts is the residence time of the trapped particles, determined using numerical tracing techniques, which is of the order of seconds and not days or weeks. These belts cannot be observed by stratospheric balloons, since their mirror fields are above the atmosphere except in correspondence of the South Atlantic Anomaly. It follows that the observed trapped particles do not belong to the various types of albedo particles reported in the past by experiments on balloons.

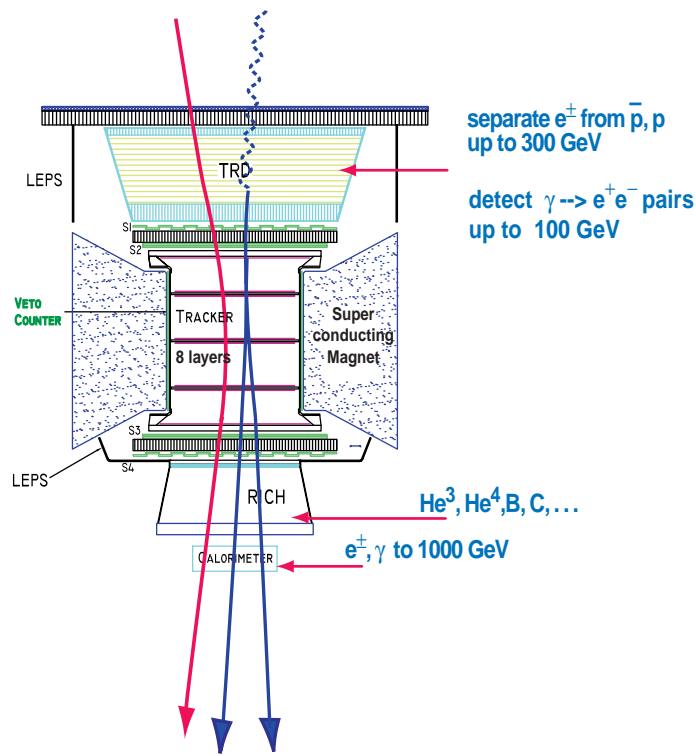

Fig. 11. Configuration of AMS on the ISS for the three years mission scheduled on the UF4 in october 2003.

In Table 2 we summarize the main features of the different type of belts identified during the last 40 years. As we can see the situation is very varied, corresponding to different filling mechanisms. Since we are dealing with continuous distributions, the reported intervals (rigidity, L, residence time) should be taken as typical order of magnitudes.

In Figure 10, we compare the structure of the AMS belts to the Van Allen belts as well as the dependence of the mirror field altitude on the longitude.

## CONCLUSIONS

The first mission of the Alpha Magnetic Spectrometer, although lasting only ten days, has been scientifically very rewarding, allowing for the first time a very detailed measurement of high energy cosmic rays outside earth atmosphere. In addition to the most accurate measurements obtained so far for the primary flux of $p, e^+, e^-, D, {}^3He$ and ${}^4He$ spectra over most of the earth surface, these results have shown the existence of a substantial second spectrum of high energy particles trapped within low altitude belts. These new belts have a very characteristic composition, dominated by positively charged particles, mainly $p$, $e^+$, and $D$. Their existence should be taken into account when calculating radiation doses for astronauts on the ISS or background rates for low orbit satellites.

AMS is currently being refurbished to be ready for a three years mission with UF4 in October 2003.

A stronger magnetic field from a superconducting magnet, $B \sim 0, 8\ T$, a fully equipped Silicon Tracker, together with three new powerful particle identification detectors, a Transition Radiation Detector, a Ring Imaging Cherenkov (RICH) detector and an Electromagnetic Calorimeter, will allow precise particle identification up to $O(TeV)$ of energy (Figure 11). The physics capabilities of AMS after three years of exposure on the ISS are summarized in Table 3. AMS will be the only large acceptance magnetic facility which will be exposed for long time in space. It will allow a measurements of the flux and composition of Cosmic Rays with an accuracy orders of magnitudes better than before. The large improvement in sensitivity given by this new instrument, will allow us to enter into a totally new domain to explore the unknown.

| Elements | Yield (or sensitivity) | (Now) | Rigidity Range (GV) | Physics |
|---|---|---|---|---|
| $e^+$ | $10^8$ | $(\sim 10^3)$ | $0.1 - 300$ | ↑ |
| $\bar{p}$ | $500000$ | $(\sim 2000)$ | $0.5 - 100$ | Dark Matter (SUSY) |
| $\gamma$ | | | $0.1 - 200$ | ↓ |
| $\bar{He}/He$ | $\frac{1}{10^9}$ | $(\frac{1}{10^6})$ | $0.5 - 1000$ | Antimatter |
| $\bar{C}/C$ | $\frac{1}{10^8}$ | $(\frac{1}{10^4})$ | $0.5 - 1000$ | CP vs GUT , EW |
| $D, H_2$ | $10^9$ | | $1.0 - 20.0$ | ↑ |
| $^3He/\ ^4He$ | $10^9$ | | $1.0 - 20.0$ | Astrophysics |
| $^{10}Be/\ ^9Be$ | $2\%$ | | $1.0 - 15.0$ | ↓ |

Table 3. Physics capabilities of AMS after three years on ISS.

## ACKNOWLEDGMENT

The timely contruction and the operation of the AMS experiment would not have been possible without the support of many national Institutes and Agencies, from US (DOE and NASA), Italy (INFN and ASI), Germany (DARA), Switzerland (SNF, ETHZ), Finland, France (IN2P3 and CNES), China (Chinese Academy of Sciences, Academia Sinica)and Taiwan (Academia Sinica, CSIST). Their support is gratefully acknowledged.

This work has been partially supported by the Italian Space Agency (ASI) under contract ARS-98/47.